# Spectral Efficiency and Optimal Medium Access Control of Random Access Systems over Large Random Spreading CDMA

Yi Sun[1]

*Abstract* – **This paper analyzes the spectral efficiency as a function of medium access control (MAC) for large random spreading CDMA random access systems that employ a linear receiver. It is shown that located at higher than the physical layer, MAC along with spreading and power allocation can effectively perform spectral efficiency maximization and near-far mitigation.**

## I. INTRODUCTION

Random access is a medium access control (MAC) protocol for multiple users to access a basestation in probabilistic manor. Random access is conventionally operated over collision channels [1]-[4]. The stability region of a finite-user slotted random access system [2][3] is identical to the information theoretical capacity region of the collision channel without feedback [1] but an infinite-user random access system over the collision channel is always unstable [4]. Random access has also been applied to the channels capable of multipacket reception [5]-[10]. It turns out [5] that multipacket reception capability can stabilize an infinite-user system. Recent research in wireless communications has paid much attention to random access [6]-[10]. The throughputs of several symmetric wireless multiaccess systems where users have the same transmission rate and power are obtained in [8]. However, the stability region and the spectral efficiency of a CDMA random access system are unknown in general. This is due partly to the fact that queues of all users are coupled by mutual interference and then the system state is a high dimensional nonhomogeneous Markov chain which is difficult to analyze. However, recent studies showed [11]-[14] that in the large system limit of random spreading (LRS) CDMA with *deterministic* access, signal to interference ratio (SIR) of linear receivers is constant. This suggests that the queues in LRS-CDMA with *random* access be decoupled and analysis be feasible [9][10]. The spectral efficiency of LRS-CDMA with deterministic access over frequency-flat fading channels is obtained in [17] with equal user power. By considering the on-off fading that is equivalent to random access, the result in [17] can be applied to CDMA random access systems with equal user power but not an arbitrary power distribution.

In this paper, for the random access system over LRS-CDMA with an arbitrary power distribution the SIR of the MMSE, decorrelator, MF is first shown to converge almost surely to a constant that captures effect of variations on traffic, spreading sequences and powers across users. The spectral efficiency is then obtained as a function of MAC, spreading, and power allocation, which therefore can jointly maximize spectral efficiency and mitigate near-far problem. In the high SIR regime MAC is more effective than power control in near-far mitigation. The results are extended to an ordinary stable system whose stability region is identical to the information theoretical capacity region.

---

[1] Yi Sun is with the Department of Electrical Engineering at the City College of City University of New York, New York, NY 10031. E-mail: ysun@ee.ccny.cuny.edu.

The obtained formulas provide a means for further study of CDMA random access systems with considerations of user collaboration, uncoded transmission, retransmission diversity, fairness, QoS, etc.

The rest of the paper is organized as follows. Section II analyzes the limit traffic load and SIR. Section III addresses the spectral efficiency, tradeoff of MAC, spreading and power, and extension to the ordinary system. All proofs are presented in the Appendix.

## II. LIMIT LINK CHANNEL

### A. Traffic load

Consider that $K$ users randomly access a basestation through a packet-switched symbol-synchronous CDMA channel. The total frequency bandwidth of the channel is $W$ Hz and the symbol period is $T_s$ seconds, and then the spreading gain equals $N = T_s W$ second Hz. We first consider a dominating system where all users have sufficiently high packet arrival rates so that they always have packets in their buffers ready for transmission. The result will be extended to an ordinary stable system where the probability for a buffer to be empty is nonzero.

Consider that user 1, the desired user, transmits a packet in a slot. In symbol period $n$ of the slot, the chip MF at the basestation outputs an $N$-dimensional vector where the index of packet slot is omitted for notation brevity,

$$\mathbf{r}(n) = \sqrt{p_1}\mathbf{s}_1(n)b_1(n) + \sum_{i=2}^{K} u_i \sqrt{p_i}\mathbf{s}_i(n)b_i(n) + \mathbf{w}(n). \qquad (1)$$

$p_1$, $\mathbf{s}_1(n)$, and $b_1(n)$ are the signal power, spreading sequence, and transmitted symbol for the desired user and $p_i$, $\mathbf{s}_i(n)$, and $b_i(n)$ for $i \geq 2$ are for interference users where $p_i$ are time-invariant. $\mathbf{w}(n) \sim N(\mathbf{0}, \sigma^2 \mathbf{I})$ are i.i.d. across symbol periods. The symbols take on real numbers with unit power $E[b_i^2(n)] = 1$ and are independent for different users. A spreading sequence can be written as $\mathbf{s}_i(n) = (s_{i1}(n), s_{i2}(n), \cdots s_{iN}(n))^T$ where the chips $s_{ij}(n)$ take on $\pm 1/\sqrt{N}$'s equiprobably and are i.i.d.

As an indication function, $u_i = 1$ if user $i$ transmits, and $u_i = 0$ otherwise. Being dependent on user's power $p_i$ and independent of other events, the transmission probability is equal to $\Pr(u_i = 1) = \theta(p_i) \in (0,1]$. Each function $\theta(p)$ that specifies transmission probabilities of all users determines a MAC scheme for the system. Though here the MAC depends on the power distribution, the results are ready to be extended to an arbitrary MAC.

Equation (1) can be written in matrix form $\mathbf{r}(n) = \sqrt{p_1}\mathbf{s}_1(n)b_1(n) + \mathbf{S}(n)\mathbf{U}\mathbf{P}^{\frac{1}{2}}\mathbf{b}(n) + \mathbf{w}(n)$ where $\mathbf{U} = \text{diag}(u_2, \ldots, u_K)$ and $\mathbf{S}(n)$, $\mathbf{P}$, and $\mathbf{b}(n)$ are specified accordingly. A linear receiver $\mathbf{l}_1(n)$ outputs an estimate of $b_1(n)$,

$$y_1(n) = \sqrt{p_1}\mathbf{l}_1^T(n)\mathbf{s}_1(n)b_1(n) + \mathbf{l}_1^T(n)\mathbf{S}(n)\mathbf{U}\mathbf{P}^{\frac{1}{2}}\mathbf{b}(n) + \mathbf{l}_1^T(n)\mathbf{w}(n).$$

By taking expectation over symbols, we obtain the power of the desired signal $P_s(n) = (\mathbf{l}_1^T(n)\mathbf{s}_1(n))^2 p_1$ and the power of interference and noise $P_{in}(n) = \sigma^2 \|\mathbf{l}_1(n)\|^2 + P_v(n)$ with interference power equal to $P_v(n) = \mathbf{l}_1^T(n)\mathbf{S}(n)\mathbf{P}\mathbf{U}\mathbf{S}^T(n)\mathbf{l}_1(n)$. The signal to interference ratio (SIR) is $P_s(n)/P_{in}(n)$, which in general varies with $n$.

We consider the LRS-CDMA channel where $K$ and $N \to \infty$ and $K/N = \alpha > 0$. $p_k$ are upper bounded and their empirical distribution function converges to $F$. Moreover, MAC $\theta(p)$ is a function such that $E[\theta(p)]$ and $E[p\theta(p)]$ exist with respect to *F*. The *traffic demand* is $\alpha$ (demanding users/s/Hz). The number of active users is $K_a = u_1 + \ldots + u_K$ and then the *traffic load* can be defined as $\rho_K = K_a/N$ active users/s/Hz or transmitted symbols/s/Hz, which varies slot by slot since $u_k$'s do so. However, as $K$ and $N$ tend to infinity, $\rho_K$ converges almost surely to a constant by the following lemma which can be proved by following the proof of Lemma 2 in the Appendix. It implies that the traffic in the LRS-CDMA *random* access system behaves like that in the LRS-CDMA *deterministic* access system [11]-[16] but the active-user number is effectively reduced by a factor of $E[\theta(p)]$.

*Lemma 1*: As $K, N \to \infty$ and $K/N = \alpha$, $\rho_K \to \alpha E[\theta(p)]$ a.s. where the expectation is taken with respect to *F*. □

*B. The limit link channel*

From the transmission of $b_1(n)$ to the estimate $y_1(n)$, there is an effective link channel. It is ready to obtain that the link channel is asymptotically memoryless Gaussian and outputs

$$y_1(n) = \sqrt{p_1\eta}\, b_1(n) + z_1(n)$$

where $z_1(n) \sim N(0,1)$ are i.i.d. and $\eta$ is the SIR when the received power is unit. The asymptotic Gaussianity of the link channel is obtained as an extension of the asymptotic Gaussianity of an LRS-CDMA *deterministic* access system in [15][16]. By Lemma 1, $K_a/N$ converges almost surely to a constant $\alpha E[\theta(p)]$ and the variation of traffic load due to the randomness of the number of active users vanishes. Therefore, the LRS-CDMA random access system in each slot is almost surely an LRS-CDMA deterministic access system where the interference of linear receiver output is Gaussian. Since spreading sequences are i.i.d. across *n*, the interferences in different symbol periods are independent. Hence, the limit link channel is Gaussian and memoryless.

In what follows, the limit SIR $\eta$ for the MMSE, decorrelator, and MF is obtained in the almost sure convergence for the sample spreading sequences. The result is applicable to the completely random sequences in the sense of deterministic convergence.

*Theorem 1*: For the LRS-CDMA random access system, as $K \to \infty$ and $K/N = \alpha$, the unit-power SIR converges almost surely to $\eta$. For the MMSE, $\eta$ uniquely solves

$$\eta^{-1} = \sigma^2 + \alpha E\left(\frac{\theta(p)p}{1+p\eta}\right). \tag{2}$$

For the decorrelator,

$$\eta = \frac{1 - \alpha E(\theta(p))}{\sigma^2} \tag{3}$$

if $\alpha E[\theta(p)] \leq 1$; otherwise, for the equal-power system with $p_i = p$, $\theta_i = \theta$, and Gaussian chips,

$$\eta = \frac{\alpha\theta - 1}{\alpha\theta\sigma^2 + (\alpha\theta - 1)^2 p}. \tag{4}$$

For the MF,

$$\eta = \frac{1}{\sigma^2 + \alpha E(\theta(p)p)}. \tag{5}$$

To each user, the SIR varies with spreading sequences and the traffic load varies slot by slot. However, the SIR in the limit system is a constant. $1/\eta$ is a notion of effective interference that captures the effect of variations in traffic, spreading sequences and user powers. It is an extension of the notion of effective interference [11] that captures only the effect of variant sequences and powers in a deterministic access system. For the MMSE, the effect of $\theta(p)$ to a user is not to simply reduce its power by a factor of $\theta(p)$ though it is for the MF. If independent of the power distribution, the MAC $\theta(p)$ for all the SIR's is effective to reduce the active-user number by a factor of $E[\theta(p)]$ as it does for the traffic load $\alpha E[\theta(p)]$. However, when the MAC is determined by the power distribution, this is not true for the MMSE and MF.

## III. SPECTRAL EFFICIENCY AND OPTIMAL MAC

### A. Spectral efficiency

Consider that a user with power $p$ transmits a packet coded at a rate $R(p)$ arbitrarily close to the link channel capacity $(1/2)\log_2(1+p\eta)$ bits/symbol (or bits per channel use). The capacity of the effective link channel including the MAC is equal to $T(p) = (\theta(p)/2)\log_2(1+p\eta)$ bits/symbol or $T(p)/(WT_s)$ bits/s/Hz. For the random access, the minimum bit energy per noise level $N_0 = 2\sigma^2$ required for reliable communication is $E_b/N_0 = \theta(p)p/(2\sigma^2 T(p))$ or

$$\frac{E_b}{N_0} = \frac{p}{\sigma^2 \log_2(1+p\eta)} \qquad (6)$$

which compared with the deterministic access is decreased due to increased $\eta$. The spectral efficiency for the entire system equals

$$C = \frac{\alpha}{2} E[\theta(p)\log_2(1+p\eta)] \qquad (7)$$

bits/s/Hz. If the MAC is independent of the power distribution, $C = (\alpha/2)E(\theta)E[\log_2(1+p\eta)]$.

Increase of transmission probabilities $\theta(p)$ can increase users' link capacities but may not increase the spectral efficiency due to the increased interference. This suggests existence of an optimal MAC that achieves the maximum spectral efficiency. In practice, a numerical optimization algorithm can be used to search the optimal $\theta(p)$. For example, consider an $M$-class system where users of class $i$ have the same power $p_i$ and therefore have the same transmission probability $\theta_i$. The expectation with respect to $F(p)$ in (7) is simplified to the summation over the $M$ classes. Optimization of MAC is then to find the optimal set of $M$ transmission probabilities $\theta_i$ such that the spectral efficiency is maximized.

Consider a two-class system where the users of class $i$ have power $p_i$, percentage of population $q_i$, and transmission probability $\theta_i$ for $i = 1, 2$. Then the spectral efficiency equals

$$C = \frac{\alpha}{2}\sum_{i=1}^{2} \theta_i q_i \log_2(1+p_i\eta)$$

where $\eta = [\sigma^2 + \alpha\theta_1 p_1 q_1/(1+p_1\eta) + \alpha\theta_2 p_2 q_2/(1+p_2\eta)]^{-1}$ for the MMSE, $\eta = [1 - \alpha(\theta_1 q_1 + \theta_2 q_2)]/\sigma^2$ for $\alpha(\theta_1 q_1 + \theta_2 q_2) \leq 1$ for the decorrelator, and $\eta = (\sigma^2 + \alpha\theta_1 p_1 q_1 + \alpha\theta_2 p_2 q_2)^{-1}$ for the MF. Fig. 1 shows spectral efficiency versus MAC $\theta(p)$ for a two-class system. The traffic demand is $\alpha = 0.95$ users/Hz/s. The users of the first and the second classes have SNR = 10 dB and 30 dB, respectively. The ratio of the user numbers of the first class to the second class is equal to $q_1/q_2 = 10$. For the MMSE, $C_{max} = 1.2$ is achieved by the MAC $(\theta_1, \theta_2) = (1,1)$, which is control-free. For the decorrelator, $C_{max} = 0.97$ is obtained by $(\theta_1, \theta_2) = (0.65, 1)$. For the MF, $C_{max} = 0.44$ is attained by the MAC $(\theta_1, \theta_2) = (1,0)$, which is unfair to the second class of users, though.

*B. Tradeoff between MAC, spreading, and power*

The spectral efficiency is a function of the number of demanding users $K$, spreading gain $N$ (or bandwidth), power $p_i$, and MAC. Optimization of these parameters can achieve the maximum spectral efficiency. $K$ and $p_i$ are dynamic in wireless systems. Adjusting MAC can timely make up the change of $K$ and $p_i$ to maintain the same spectral efficiency, which is the unique advantage of random access systems over deterministic systems. In the case when the MAC is independent of power distribution, if $K$ is changed by a factor of $\gamma$, then adjusting the MAC to change $E(\theta)$ by a factor of $1/\gamma$ can maintain the same spectral efficiency.

To solve the near-far problem in wireless communication, power control is usually performed so that all users have about the same link capacity. As can be seen from $\eta$, reduction of transmission probability is equivalent to reduction of their interference to other users. Moreover, the formula of $T(p)$ shows that increase of transmission probability can effectively increase their link capacities. It is clear that a proper MAC, even without power control, can achieve the goal of unifying link capacities. In practice, MAC and power control can be jointly applied. However, sensitivities of link capacity to transmission probability and to transmission power are different in different SIR regimes. In the low SIR regime such that $\eta p \ll 1$, the link capacity is $T(p) \cong \theta p \eta / (2 \ln 2)$. Both MAC and power control can effectively change the link capacity. On the other hand, in the high SIR regime such that $\eta p \gg 1$, the link capacity is $T(p) \cong \theta \ln(p\eta)/(2\ln 2)$. MAC is more effective than power control in change of link capacity.

*C. Extension to a stable system*

Now we extend the result from the dominating system to an ordinary stable system where the arrival rate is less than the service rate so that a user's buffer is empty with a nonzero probability.

A user with power $p$ has a queue of packets in its buffer where new packet arrival rate is $\lambda(p)$ packets/slot. The packet at the front end is transmitted with probability $\theta(p)$. Since packets are coded with an achievable rate, a transmitted packet is successfully decoded with probability one. Thus, the service rate of the queue is equal to $\theta(p)$. We shall address, in the conventional terms of random access, the stability, probability distribution of queue length, and packet delay. This implies that the packet size is finite, seemly contradictory to the fact that the size of packets coded with an achievable rate shall tend to infinity. However, the result is in the sense that if the packet size is sufficiently large, the system shall behave sufficiently close to the result as the packet size tends to infinity.

A queue is stable if its length is finite with probability one. If almost every queue in the system is stable as $K \to \infty$, then the system is stable. The stability region is the collection of sets of all users' arrival rates with which the system is stable. In the LRS-CDMA system, the traffic load in each slot is equal to the constant $E[\theta(p)]$ and the interference to a user is the constant $1/\eta$ both almost surely. Hence, the queue's behavior depends only on its arrival and service rates. It is clear that the queue is stable if $\lambda(p) < \theta(p)$, and unstable if $\lambda(p) > \theta(p)$. Moreover, it is notable that when $\lambda(p) < \theta(p)$, the probability that the queue is empty is nonzero and then the interference power to other users is less than that in the dominating system. Hence, $\lambda(p) < \theta(p)$ for all $p$ specifies the stability region. Furthermore, the information arrival rate satisfies $r(p) = \lambda(p)R(p)/(WT_s) < \theta(p)\log(1+p\eta)/(2WT_s) = T(p)/(WT_s)$ bits/s/Hz. Therefore, the stability region is identical to the information theoretical capacity region. The spectral efficiency (7) is the supreme of sum rate over the stability region.

For a stable queue, the probability distribution $q_m$ that the queue has $m$ packets in a slot is stationary. The rate of packets flowing from state $m+1$ to $m$ is equal to the rate from $m$ to $m+1$. That is, $(1-\lambda(p))\theta(p)q_1 = \lambda(p)q_0$ and $(1-\lambda(p))\theta(p)q_m = \lambda(p)(1-\theta(p))q_{m-1}$ for $m \geq 2$. Since $\sum_{m=0}^{\infty} q_m = 1$, we obtain $q_0 = 1 - \lambda(p)/\theta(p)$ $q_m = q_0 \beta^{m-1}(p)\lambda(p)/[(1-\lambda(p))\theta(p)]$ for $m \geq 1$ where $\beta(p) = \lambda(p)(1-\theta(p))/[(1-\lambda(p))\theta(p)]$. The probability that the queue is empty is $q_0 > 0$ due to $\lambda(p) < \theta(p)$. The packet delay is the average number of slots from the slot of a packet arrival at the queue to the slot of its service. Since a packet is coded at an achievable rate, the packet delay is equal to the average length of the queue, $D = \sum_{m=1}^{\infty} m q_m = \lambda(p)(1-\lambda(p))/(\theta(p)-\lambda(p))$. It is clear that in each slot a user of power $p$ transmits a packet with probability $(1-q_0)\theta(p) = \lambda(p)$, which equals the arrival rate. This is due to the fact that in a stable system, the departure rate is equal to the arrival rate. Hence, the MAC can serve to control $\lambda(p)$. Then, replacing $\theta(p)$ by $\lambda(p)$, all the formulas and results in the preceding subsections are applicable to the stable system.

## APPENDIX

*Lemma 2*: Let $\{p_i\}$ be an upper bounded sequence and its empirical distribution function converges to $F$, that is $\lim_{K \to \infty}(1/K)\sum_{i=1}^{K} I(p_i \leq x) = F(x)$ and $u_i$'s be independent with $\Pr(u_i = 1) = \theta(p_i)$ and $\Pr(u_i = 0) = 1 - \theta(p_i)$. As $K \to \infty$, $(1/K)\sum_{i=1}^{K} \theta(p_i) \to E(\theta(p))$ where $E$ is with respect to $F$. Then the empirical distribution function of $\{p_i u_i\}$, $H_K(x) = (1/K)\sum_{i=1}^{K} I(p_i u_i \leq x)$, converges almost surely to $H(x) = 1 - E(\theta(p)) + \int_0^x \theta(p)dF(p)$.

*Proof*: Since $I(u_i p_i \leq x) = I(u_i = 0) + I(u_i = 1)I(p_i \leq x)$, $H_K(x) = (1/K)\sum_{i=1}^{K} I(u_i = 0) + V_K(x)$ where

$$V_K(x) = \frac{1}{K} \sum_{i=1}^{K} I(u_i = 1)I(p_i \leq x). \tag{8}$$

The mean of $V_K(x)$ is $E[V_K(x)] = (1/K)\sum_{i=1}^{K} \theta(p_i)I(p_i \leq x)$, which converges to $\int_0^x \theta(p)dF(p)$. Moreover,

$$E[(V_K(x) - E(V_K(x)))^4] = \frac{1}{K^4} \sum_{i=1}^{K} E[(I(u_i = 1) - \theta(p_i))^4]I(p_i \leq x)$$

$$+\frac{3}{K^4}\sum_{i=1}^{K}\sum_{j=1,j\neq i}^{K}E[(I(u_i=1)-\theta(p_i))^2]E[(I(u_j=1)-\theta(p_j))^2]I(p_i\leq x)I(p_j\leq x)$$

$$\leq \frac{1}{K^3}+\frac{3(K-1)}{K^3}=O(1/K^2). \tag{9}$$

By Chebyshev inequality, for each $\varepsilon > 0$, $\Pr\{|V_K(x)-E(V_K(x))|>\varepsilon\}\leq c/(\varepsilon^4 K^2)$, and therefore it follows from the Borel-Cantelli lemma that $V_K(x)-E(V_K(x))\to 0$ a.s. and then $V_K(x)\to \int_0^x \theta(p)dF(p)$ a.s. Following the similar lines, we can show that $H_K(x) - V_K(x) \to 1 - E[\theta(p)]$ a.s. □

*Proof of Theorem 1*:

By ignoring a factor of positive constant, the MMSE receiver is $\mathbf{l}_1(n)=\mathbf{M}^{-1}(n)\mathbf{s}_1(n)$ where $M(n)=\sigma^2\mathbf{I}+\mathbf{S}(n)\mathbf{P}\mathbf{U}\mathbf{S}^T(n)$. The SIR equals $P_s(n)/P_{in}(n)=p_1\mathbf{s}_1^T(n)\mathbf{M}^{-1}(n)\mathbf{s}_1(n)$. Let $\Lambda(n)=\mathrm{diag}(\lambda_1,\ldots,\lambda_N)$ with $\lambda_i$ be the eigenvalues of $\mathbf{S}(n)\mathbf{P}\mathbf{U}\mathbf{S}^T(n)$ and $\mathbf{Z}(n)$ be the corresponding matrix of eigenvectors. Denote by $G_N$ the empirical distribution function of the eigenvalues. By [11] (Theorem 4.1), $G_N$ converges almost surely to $G$ whose Stieltjes transform equals $m(z)=\{-z+\alpha\int_0^\infty \{t/[1+tm(z)]\}dH(t)\}^{-1}$ where $H$ is the limit empirical distribution function of $p_iu_i$'s. It follows from Lemma 2 that

$$m(z)=\int_0^\infty \frac{1}{\lambda-z}dG(\lambda) \tag{10}$$

$$=\frac{1}{-z+\alpha\int_0^\infty \{\theta(p)p/[1+pm(z)]\}dF(p)}. \tag{11}$$

For any $\mathbf{U}$ and $N$, $\mathbf{S}(n)\mathbf{P}\mathbf{U}\mathbf{S}^T(n)$ is nonnegative definite and therefore the spectral radius of $\mathbf{M}^{-1}(n)$ is upper bounded by $1/\sigma^2$. By [15] (Lemma B.1),

$$E\left\{\left|\mathbf{s}_1^T(n)M^{-1}(n)\mathbf{s}_1(n)-\frac{1}{N}\mathrm{tr}(M^{-1}(n))\right|^6\right\}\leq \frac{c}{N^3}. \tag{12}$$

It follows from the Chebyshev inequality and the Borel-Cantelli lemma that $\mathbf{s}_1^T(n)M^{-1}(n)\mathbf{s}_1(n)-(1/N)\mathrm{tr}(M^{-1}(n))\to 0$ a.s. Since $(1/N)\mathrm{tr}(M^{-1}(n))=(1/N)\mathrm{tr}[\mathbf{Z}(n)(\sigma^2\mathbf{I}+\Lambda(n))^{-1}\mathbf{Z}^T(n)]=(1/N)\sum_{i=1}^N 1/(\sigma^2+\lambda_i)=\int_0^\infty 1/(\sigma^2+\lambda)dG_N(\lambda) \to \int_0^\infty 1/(\sigma^2+\lambda)dG(\lambda)$ a.s., we obtain

$$\mathbf{s}_1^T(n)M^{-1}(n)\mathbf{s}_1(n)\to \eta = \int_0^\infty \frac{1}{\sigma^2+\lambda}dG(\lambda) \quad \text{a.s.} \tag{13}$$

Letting $z\to -\sigma^2$ in (10), the continuity of $m(z)$ yields $m(z)\to \eta$, which leads (11) to $\eta$.

The decorrelator $\mathbf{l}_1(n)$ of user 1 is the first column vector of $[(\mathbf{s}_1(n),\mathbf{S}(n)\mathbf{U})^+]^T$ where $+$ denotes pseudoinverse. If $\mathbf{s}_1(n)$ is not in the space spanned by $\mathbf{S}(n)\mathbf{U}$, then the decorrelator cancels out the interference signal and

$\mathbf{l}_1(n) = [\mathbf{I} - \mathbf{S}(n)\mathbf{U}(\mathbf{S}(n)\mathbf{U})^+]\mathbf{s}_1(n)$. The SIR is $P_s(n)/P_{in}(n) = p_1 \mathbf{s}_1^T(n)[\mathbf{I} - \mathbf{S}(n)\mathbf{U}(\mathbf{S}(n)\mathbf{U})^+]\mathbf{s}_1(n)/\sigma^2$. Let $\mathbf{S}(n)\mathbf{U}$ be expressed in the singular value decomposition as $\mathbf{S}(n)\mathbf{U} = \sum_{i=1}^{K'_a(n)} \lambda_i \mathbf{z}_i \mathbf{v}_i^T$ where $\lambda_i$'s are $K'_a(n) \leq K_a(n)$ nonzero singular values. Its pseudo-inverse is $(\mathbf{S}(n)\mathbf{U})^+ = \sum_{i=1}^{K'_a(n)} \lambda_i^{-1} \mathbf{v}_i \mathbf{z}_i^T$ and then $\mathbf{I} - \mathbf{S}(n)\mathbf{U}(\mathbf{S}(n)\mathbf{U})^+ = \mathbf{I} - \sum_{i=1}^{K'_a(n)} \mathbf{z}_i \mathbf{z}_i^T$. The spectral radius of $\mathbf{I} - \mathbf{S}(n)\mathbf{U}(\mathbf{S}(n)\mathbf{U})^+$ is upper bounded by one. It follows from [15] (Lemma B.1) that $E\left\{\left|\mathbf{s}_1^T(n)[\mathbf{I} - \mathbf{S}(n)\mathbf{U}(\mathbf{S}(n)\mathbf{U})^+]\mathbf{s}_1(n) - (1/N)\mathrm{tr}[\mathbf{I} - \mathbf{S}(n)\mathbf{U}(\mathbf{S}(n)\mathbf{U})^+]\right|^6\right\} \leq c/N^3$. By the Chebyshev inequality and the Borel-Cantelli lemma, $\mathbf{s}_1^T(n)[\mathbf{I} - \mathbf{S}(n)\mathbf{U}(\mathbf{S}(n)\mathbf{U})^+]\mathbf{s}_1(n) - \frac{1}{N}\mathrm{tr}[\mathbf{I} - \mathbf{S}(n)\mathbf{U}(\mathbf{S}(n)\mathbf{U})^+] \to 0$ a.s. By Lemma 1, $K_a(n)/N \to \alpha E(\theta(p))$ a.s. Suppose $K_a(n)/N < 1$. By [18] (Proposition 2.1), the least singular value is almost surely greater than zero, and therefore $\mathbf{S}(n)\mathbf{U}(n)$ has a full column rank (i.e., $K'_a(n) = K_a(n)$) almost surely. It follows from Lemma 1 that $K'_a(n)/N \to \alpha E(\theta(p))$ a.s. and thus $(1/N)tr[\mathbf{I} - \mathbf{S}(n)\mathbf{U}(\mathbf{S}(n)\mathbf{U})^+] \to 1 - \alpha E(\theta(p))$ a.s. The second part is obtained by following the lines in [13] [14] and noticing that the traffic load is $K_a(n)/N \to \alpha E(\theta(p))$ a.s.

The MF is $\mathbf{l}_1(n) = \mathbf{s}_1(n)$. The signal power equals $P_s(n) = p_1$ and the interference power equals $P_v(n) = \mathbf{s}_1^T(n)\mathbf{S}(n)\mathbf{PU}\mathbf{S}^T(n)\mathbf{s}_1(n)$. It is sufficient to show that $P_v(n) \to \alpha E[\theta(p)p]$ a.s. By [18] (Proposition 2.1), $\|\mathbf{S}^T(n)\mathbf{S}(n)\| \leq (1+\sqrt{\alpha})^2$ a.s., and therefore $\|\mathbf{S}(n)\mathbf{PU}\mathbf{S}^T(n)\| \leq \|\mathbf{S}(n)\| \|\mathbf{PU}\| \|\mathbf{S}^T(n)\| \leq (1+\sqrt{\alpha})^2 \max_{2 \leq k \leq K}\{p_i\}$, which is upper bounded. Then the spectral radius of $\mathbf{S}(n)\mathbf{PU}\mathbf{S}^T(n)$ is upper bounded in $N$. It follows from [15] (Lemma B.1) that $E\left\{\left|P_v(n) - (1/N)\mathrm{tr}(\mathbf{S}(n)\mathbf{PU}\mathbf{S}^T(n))\right|^6\right\} \leq c/N^3$. In terms of the Chebyshev inequality and the Borel-Cantelli lemma, $P_v(n) - (1/N)\mathrm{tr}(\mathbf{S}(n)\mathbf{PU}\mathbf{S}^T(n)) \to 0$ a.s. Moreover, $(1/N)\mathrm{tr}(\mathbf{S}(n)\mathbf{PU}\mathbf{S}^T(n)) = (1/N)\mathrm{tr}(\mathbf{PU}\mathbf{S}^T(n)\mathbf{S}(n))$ $= (1/N)\sum_{k=2}^{K} p_k u_k [\mathbf{S}^T(n)\mathbf{S}(n)]_{kk} = (1/N)\sum_{k=2}^{K} p_k u_k$, which by Lemma 2 converges almost surely to

$$\alpha \int_0^\infty p\, dH(p) = \alpha \int_0^\infty \theta(p) p\, dF(p). \tag{14}$$

This completes the proof. □

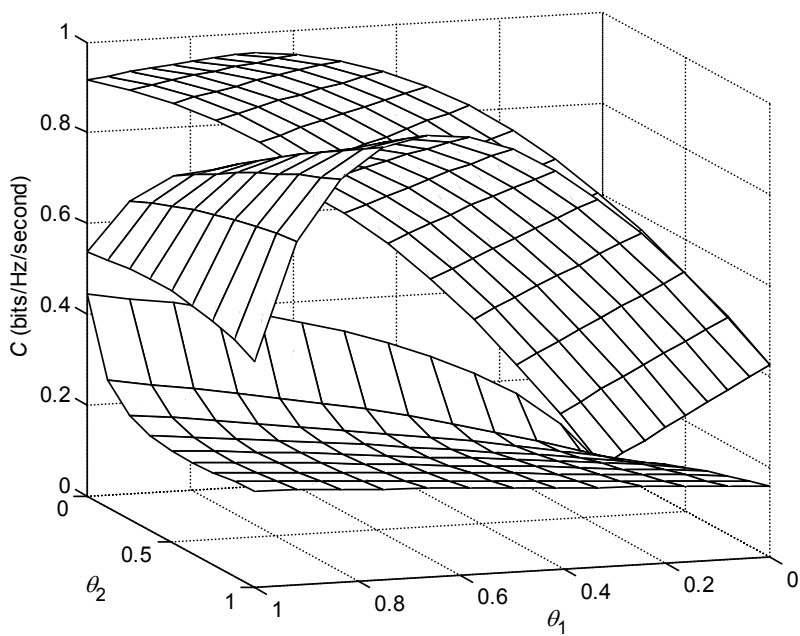

Fig. 1. Spectral efficiency versus MAC ($\theta_1, \theta_2$) for a two-class system. From up to down are for the MMSE, decorrelator, and MF, respectively.